
\input harvmac
\def\{{\lbrace}
\def\}{\rbrace}
\def\({\lbrack}
\def\){\rbrack}
\def\half{{1\over 2}}
\def\p{\partial}
\def\pab{\bar{\p}}
\def\ph{\phantom{-}}
\def\sTr{{\rm sTr}}
\def\A{{\cal A}}
\def\F{{\cal F}}
\def\Z{{\cal Z}}

\def\zb{\bar{z}}
\def\th{\theta}
\def\tb{\bar{\th}}
\def\dab{\bar{D}}

\def\smat#1#2#3#4{\left(\matrix{#1&\!\!\vrule\!\!&#2\cr
                                \noalign{\smallskip\hrule\smallskip}
                                #3&\!\!\vrule\!\!&#4\cr}\right)}

\font\twelvebf=cmbx12
\nopagenumbers

\ref\walgebra{P. Mathieu, {\it Phys.Lett.} {\bf B208} (1988) 101; \hfill
\break
I. Bakas, {\it Commun.Math.Phys.} {\bf 123} (1989) 627; \hfill
\break
P. Di Francesco, C. Itzykson and J.-B. Zuber,
{\it Commun.Math.Phys.} {\bf 140} (1991) 543.}
\ref\hn{K. Hiutu and D. Nemeschansky,
{\it Mod.Phys.Lett.} {\bf A6} (1991) 3179; \hfill\break
T. Inami and H. Kanno, {\it J.Phys.} {\bf A25} (1992) 3729.}
\ref\gt{F. Gieres and S. Theisen,
``Superconformally Covariant
Operators and Super $W$-Algebras", preprint MPI-Ph/92-66.}
\ref\bfk{A. Bilal, V.V. Fock and I.I. Kogan,
{\it Nucl.Phys.} {\bf B359} (1991) 635; \hfill
\break
A. Das, W.-J. Huang and S. Roy, {\it Int.Mod.J.Phys.} {\bf A7} (1992) 3447.}
\ref\japan{S. Komata, K. Mohri and H. Nohara,
{\it Nucl.Phys.} {\bf B359} (1991) 168.}
\ref\cuba{J.M. Figueroa-O'Farrill and E. Ramos, {\it Commun.Math.Phys.}
{\bf 145} (1992) 43; {\it Nucl. Phys.} {\bf B368} (1992) 361.}
\ref\drs{F. Delduc, E. Ragoucy and P. Sorba,
{\it Commun.Math.Phys.} {\bf 146} (1992) 403.}
\ref\corn{J.F.Cornwell, ``Group Theory and Physics",
Volume 3 (Academic, London 1990).}
\ref\cco{F. Gieres, {\it Int.J.Mod.Phys} {\bf A8} (1993) 1.}
\ref\balog{J. Balog, L. Feher, P. Forgacs, L. O'Raifeartaigh and A. Wipf,
{\it Ann.Phys.} {\bf 203} (1990) 76.}
\ref\dg{F. Delduc and F. Gieres, {\it Int.J.Mod.Phys.} {\bf A7} (1992) 1685.}
\ref\blum{J.M. Figueroa-O'Farrill and S. Schrans, {\it Phys.Lett.}
{\bf B245} (1990) 471;\hfill\break
R. Blumenhagen, {\it Nucl. Phys.} {\bf B381} (1992) 641.}
\ref\gates{S.J. Gates, JR., M.T. Grisaru, M. Rocek and W. Siegel,
``Superspace" (Benjamin/Cummings, New York 1993); \hfill
\break
F. Gieres: ``Geometry of Supersymmetric Gauge Theories",
Lecture Notes in Physics 302 (Springer, Berlin 1988).}
\vfill\eject

\rightline{LMU-TPW 93-07}
\rightline{March 1993}
\vskip .7in

\centerline{\twelvebf  Classical
$N=1$ and $N=2$ super W-algebras}
\vskip.3truecm
\centerline{\twelvebf  from a zero-curvature condition}
\vskip1cm
\centerline{Fran\c cois Gieres$^*$ and Stefan Theisen$^{**}$}
\medskip
\centerline{$^*${\it Laboratoire de
Math\'ematiques Appliqu\'ees}\footnote{$^{\S}$}
{Unit\'e de recherche
URA 1501 du CNRS.}}
\centerline{{\it Universit\'e B. Pascal (Clermont-Ferrand II),
F - 63177 Aubi\`ere
C\'edex}}
\medskip
\centerline{$^{**}${\it Sektion Physik der Universit\"at M\"unchen}}
\centerline{{\it Theresienstra\ss e 37, D - 8000 M\"unchen 2, FRG}}
\vskip .5in
\centerline{{\bf Abstract}}
\smallskip
\noindent
Starting from superdifferential operators in an $N=1$
superfield formulation, we present a systematic
prescription for the derivation of
classical $N=1$ and $N=2$ super W-algebras by imposing
a zero-curvature condition on the connection of the
corresponding first order system.
We illustrate the procedure on the
first non-trivial example (beyond the $N=1$ superconformal
algebra) and also comment on the relation with the Gelfand-Dickey
construction of $W$-algebras.

\vfill\eject

\footline={\hss\tenrm\folio\hss}
\pageno=1
\noindent {\bf 1. Introduction}

It is well known how to derive classical W-algebras from
(pseudo) differential operators by virtue of
the second Gelfand-Dickey
Hamiltonian structure \walgebra.
This procedure has recently been generalized to
the supersymmetric case \hn~
(see also \gt).
Here we will generalize another construction
which starts from the same differential operators. This method
amounts to describing a $n$-th order operator
in terms of a system of
first order operators and imposing a zero-curvature condition
on the associated connection (and its anti-analytic partner).
This has
previously been done for the case of
$sl(n)$(-valued connections) \bfk.
Here, we
will provide the necessary algebraic tools for the
$N=2$ super W-algebras (SWA's) corresponding to
the super Lie algebras $sl(n+1|n)$ and
for the $N=1$ SWA's corresponding to
$osp(2m\pm1|2m)$. The connection between the super W- and
the super Lie algebras \japan \cuba \drs~
will become transparent in the
discussion. The case of $osp(3|2)$ will be presented
in detail in section 6 where we also exhibit the
relation between the zero-curvature construction
of (super) $W$-algebras and the Gelfand-Dickey approach.
In Appendix A we develop the superspace formulation
of gauge theories based on graded Lie algebras in order
to give a precise geometric meaning to the zero-curvature
condition which represents the starting point of our analysis.

\noindent {\bf 2. Super differential operators}

To start with,
we consider the most general
superconformally covariant differential
operator of order $2n+1$ acting on superconformal (primary) fields.
As shown in reference \gt,
it may be cast into the form
\eqn\genop{
{\cal L}^{(n)}=D^{2n+1}+a_2 D^{2n-1}+\dots +a_{2n+1} \, ,}
where $D = \p_{\theta}
+ \theta \p$ and $\p =\p _z$. Under a superconformal
change of the local coordinates $(z,\theta)$,
the coefficient functions $a_k (z, \theta)$
transform in such a way that
$$
{\cal L}^{(n)}:~{\cal F}_{-n}\to{\cal F}_{n+1} \, ,
$$ where
${\cal F}_k$ represents the space of
primary fields of weight ${k\over 2}$. The
$a_k$ are even (odd) for $k$ even (odd).
More specifically, $a_2$ belongs to ${\cal F}_2$ and
$a_3-\half Da_2$ transforms like a
superprojective connection (i.e. like the energy momentum tensor).
The remaining coefficients $a_{k}$ ($k\geq4$)
transform in a more complicated way.
However, one can
express the operator ${\cal L}^{(n)}$
entirely in terms of fields $V_2,V_3,\dots ,V_{2n+1}$
where $V_3$ transforms like the energy momentum tensor and where
all the other $V_k$ are primary fields of weight ${k\over 2}$.
The relation between the $a_k$ and the $V_k$ is most easily
obtained by rewriting the superdifferential equation
\eqn\start{
{\cal L}^{(n)}f=0 \qquad \quad
(\, f \in {\cal F}_{-n}\, )\, ,}
which is of order $2n+1$, as a system of
first order differential equations.
For the form \genop~ of the operator ${\cal L}^{(n)}$, equation
\start~ is equivalent to
\eqn\nablaprime{
\nabla^\prime\vec F^\prime=0 \, , }
where
\eqn\f{
\nabla^\prime \equiv D-\A^\prime(a_k)  \equiv
\pmatrix{D&a_2&a_3&\dots& a_{2n+1}\cr
         -1&\ddots&&&\cr
           &\ddots&\ddots&&\cr
           &&\ddots&\ddots&\cr
           &&&-1&D\cr}\,,\qquad
\vec F^\prime=(D^{2n} f,D^{2n-1}f,\dots,f)^T \,.}
(Here and below the non-displayed matrix elements are
assumed to be zero.)
On the other hand, if
${\cal L}^{(n)}$ is parametrized in terms
of the fields $V_k$, equation \start~ is equivalent to
\eqn\del{
\nabla \vec F=0\, , }
where
\eqn\hwg{
\nabla\equiv D-\A(V_k)  \equiv
\pmatrix{D&&&&\cr
         -1&\ddots&&V's&\cr
           &\ddots&\ddots&&\cr
           &&\ddots&\ddots&\cr
           &&&-1&D\cr}\,,\qquad
\vec F=(f_1,\dots,f_{2n+1})^T \, , \,f_{2n+1} \equiv f \,.}
The upper triangular part
in $\A(V_k)$ denoted by `$V's$' will be made explicit
in a concrete example in section 6. The superfield
$\A$ takes its values in the super Lie algebra $sl(n+1|n)$
equipped with the diagonal grading, which means the following
\gt.
The graded algebra
$sl(n+1|n)$ consists
of $(2n+1)\times(2n+1)$ supermatrices
with vanishing supertrace (e.g. see \corn).
The standard grading is
defined in terms of even and odd blocks: $M=\smat{A}{B}{C}{D}$
with $A_{(n+1)\times(n+1)}$ and $D_{n\times n}$ even and
$B$ and $C$ odd. Supertracelessness means
$0=\sTr \, M=\Tr \, A-\Tr \, D$. However, from
the expression for ${\cal A}$ we see that the natural
grading for our discussion is not the standard one, but rather
a grading by diagonals, i.e. we have even and odd diagonals.
The supertrace is then given by $\sTr \, M=\sum_i(-)^i M_{ii}$ and the graded
commutator by
$$
[ M , N \} _{ik} = \sum_j \Bigl( M_{ij} N_{jk} - (-)^{(i+j)(j+k)}
N_{ij} M_{jk} \Bigr)
\, .
$$

It is straightforward (at least
in principle) to express the $V_k$ in terms
of the $a_k$ and vice versa. In fact, by virtue of
the equation $\nabla\vec F=0$
we can express $\vec F$ in terms of $\vec F^\prime$ as
$\vec F=X\vec F^\prime$ where $X$ belongs to the supergroup
$SL(n+1|n)$.
Then $\nabla^\prime=\widehat X^{-1}
\nabla X$ where the hat means that the signs of all odd generators in
$X$ have been reversed \footnote{$^*$}{If $X = {\rm exp} \, x$ with
$x \in sl(n+1 |n)$, then $\widehat X = {\rm exp} \, \widehat x$.}
\drs. From
now on, we will always work with $\nabla$ rather than $\nabla^{\prime}$.
Moreover, we will always assume that the coefficient functions $V_k$
are supersmooth rather than superholomorphic. (This assumption does not affect
the transformation properties \cco, since the operator ${\cal L}^{(n)}$
only contains derivatives with respect to $z$ and $\th$.)

\noindent {\bf 3. Zero-curvature condition}

In order to impose a zero-curvature condition on the connection,
we need to pair the
connection component $\A_{\th} \equiv \A$ with components
$\A_{\tb} \equiv \bar{\A}$ and
$\A_z , \A_{\zb}$.
These smooth superfields all take their values in $sl(n+1|n)$ or
a subalgebra thereof. The component $\A$ will always be assumed to be in
the so-called {\it
highest weight gauge} form \hwg~ \balog \drs.
We will see that choosing $\bar\A\in sl(n+1|n)$ will lead to $N=2$
SWA's and choosing $\bar\A\in osp(2m\pm1|2m) \subset sl(n+1|n)$
with $4m\pm 1 = 2n+1$ will lead to
$N=1$ SWA's.
As usual in supergauge theories, the spatial components
$\A_z$ and $\A _{\bar z}$ of the connection depend on the other
components, $\A$ and $\bar \A$,
by virtue of redefinition constraints
(see Appendix A) and therefore we will not consider them
in the sequel.

If we simultanously
consider the two first-order systems $\nabla \vec F = 0$
and $\bar{\nabla} \vec F =0 $, i.e.
\eqn\fos{
D\vec F=\A\vec F\,,\quad
         \bar D\vec F=\bar\A\vec F\, , }
we have to require their compatibility, i.e. $\{D,\bar D\} \vec F =0$.
In order to investigate the implications of this relation, we decompose
$\A \equiv \A_{\th}$ and $\bar \A \equiv \A_{\tb}$
along the even and odd basis elements
($t_a$ and $t_{\alpha}$)
of the super Lie algebra under consideration:
\eqn\decomp{
\A = \A ^a t_a + \A ^{\alpha} t_{\alpha} \qquad , \qquad
\bar{\A} = \bar{\A} ^a t_a + \bar{\A} ^{\alpha} t_{\alpha} \, .}
The superfields $\A^a$ and $\bar\A ^a$ have odd parity
while $\A^{\alpha}$ and $ \bar{\A}^{\alpha}$
have even parity. Since $\dab$ is odd, we get
\eqn\odd{\eqalign{
\dab (D \vec F) =
\dab (\A \vec F) = & (\dab \A )\vec F + (
- \A ^a t_a + \A ^{\alpha} t_{\alpha} )\dab \vec F \cr
= &\;
(\dab \A ) \vec F - \widehat{\A}  \dab \vec F \cr
= & \; ( \dab \A - \widehat{\A} \bar{\A} )\vec F \, . }}
Thus
the integrability condition
$\{D,\bar D\} \vec F =0$ is equivalent to the relation
\eqn\zerocurv{
D\bar\A+\bar D\A-\widehat\A\bar\A
-\widehat{\bar\A}\A =0 \, . }
This equation represents
a zero-curvature condition for the connection since it
corresponds to the vanishing of the curvature
component ${\cal F}_{\th \tb}$
(see Appendix A).

\goodbreak
\noindent {\bf 4. The superalgebra $sl(n+1|n)$}

In this section we briefly summarize some results which
are relevant for our subsequent discussions.
The Cartan matrix $(a_{ij})$ of $sl(n+1|n)$ has the following
non-vanishing elements \japan \drs:
\eqn\cartan{
a_{i,i+1}=a_{i+1,i}=(-)^{i+1} \qquad  \quad
(\ i=1,\dots,2n  \ ) \, .}
A basis for the Chevalley generators with diagonal grading
and with all simple roots fermionic is given by \gt
\eqn\chevalley{\eqalign{
h_i&=(-)^{i+1}\Bigl(E_{i,i}+E_{i+1,i+1}\Bigr)\cr
e_i&=(-)^{i+1}E_{i,i+1}\cr
f_i&=E_{i+1,i}\, ,}}
where $(E_{ij})_{kl} = \delta_{ik} \delta_{jl}$.
A general element of $sl(n+1|n)$ with diagonal grading is then
simply a $(2n+1)\times(2n+1)$ matrix of numbers together with
an even (odd) grading for the even (odd) diagonals,
the only restriction being the vanishing of the supertrace.

The principal embedding $osp(1|2)_{\rm pal}\subset sl(n+1|n)$
is given by \drs \gt
\eqn\sppal{\eqalign{
J_-&=\sum_i f_i={\rm diag}_{-1}(1,\dots,1)\cr
J_+&=\sum_{ij}a^{ij}e_i={\rm diag}_{+1}(n,-1,n-1,-2,\dots,-n)\cr
H&=\{J_+,J_-\}={\rm diag}_0(n,n-1,\dots,-n) \, ,}}
where we count the main diagonal as the 0-th one.
We have the commutation relations
$$
\(H,e_i\)=e_i \qquad , \qquad \(H,f_i\)=-f_i \, .
$$
The highest weight generators of $osp(1|2)_{\rm pal}$
are defined as those $M_k\in sl(n+1|n)$ which satisfy
\eqn\high{
\(H,M_k\)=kM_k \qquad {\rm and} \qquad
\(J_+,M_k\}=0 \, .}
For the solutions of these conditions one finds $M_k=M_1^k$ with
\eqn\hweights{M_1={\rm diag}_{+1}(n,1,n-1,2,\dots,n)
\qquad \quad (\ M_1^{2n+1}=0\ ) \,.}
Henceforth \gt,
the super Lie algebra
$sl(n+1|n)$ is associated with a connection
(in highest weight gauge)
\eqn\slncon{-{\cal A}_{N=2}=-J_-+\sum_{k=1}^{2n}V_{k+1}M_k \, ,}
where $V_3$ is a superprojective connection and
$V_k \in \F_k$ for $k=2,4,\dots 2n+1$.
The first order system $D- \A_{N=2}$ is equivalent to a
covariant superdifferential
operator of
order $2n+1$
of the form \genop~
without any restrictions on its
coefficients.

\goodbreak
\noindent {\bf 5. The superalgebras $osp(2m\pm 1|2m)$}

For the discussion of $N=1$ SWA's,
the graded algebras $osp(2m\pm1|2m)\subset sl(n+1|n)$ with
$4m\pm1=2n+1$ are relevant \japan \drs. In order to specify the
zero-curvature condition to this case, we need a general characterization
of the elements of these algebras in diagonal grading.
This will be provided in the following.
Some more details concerning the
diagonal grading representation of
$osp(2m\pm1|2m)$ are presented in
Appendix B.

In a basis of the root space with only fermionic simple roots,
the algebras
$osp(2m\pm1|2m)$ are characterized by Cartan matrices
$a_{ij}$ whose only non-zero elements are \japan~
\eqn\elie{
a_{11}=1 \quad , \quad \,a_{i,i+1}=a_{i+1,i}=(-)^i \, ,}
where $i=1,\dots, 2m$
for $osp(2m+1|2m)$ and $i=1,\dots,2m-1$ for $osp(2m-1|2m)$.
The Chevalley generators can be chosen as follows:
\vskip.5cm
\noindent $osp(2m+1|2m)$:
\eqn\chevalleyp{\eqalign{
h_i&=(-)^{i+1}\Bigl(E_{2m+1-i,2m+1-i}-E_{2m+1+i,2m+1+i}
    +E_{2m+2-i,2m+2-i}-E_{2m+i,2m+i}\Bigr)\cr
e_i&=(-)^i\Bigl(E_{2m+i,2m+1+i}-E_{2m+1-i,2m+2-i}\Bigr)\cr
f_i&=E_{2m+1+i,2m+i}+E_{2m+2-i,2m+1-i}}}
\vskip.5cm
\noindent $osp(2m-1|2m)$:
\eqn\chevalleym{\eqalign{
h_i&=(-)^{i+1}\Bigl(E_{2m-i,2m-i}-E_{2m+i,2m+i}
    +E_{2m+1-i,2m+1-i}-E_{2m-1+i,2m-1+i}\Bigr)\cr
e_i&=(-)^i\Bigl(E_{2m-1+i,2m+i}-E_{2m-i,2m+1-i}\Bigr)\cr
f_i&=E_{2m+i,2m-1+i}+E_{2m+1-i,2m-i}\ .}}
They satisfy the algebra
\eqn\algebra{\eqalign{
       \(h_i,e_j\)&=\ph a_{ij}e_j\cr
       \(h_i,f_j\)&=-a_{ij}f_j\cr
       \{e_i,f_j\}&=\delta_{ij} h_j\, .}}

A general $osp(2m\pm1|2m)$ element in diagonal grading is then
a $(4m\pm1)\times(4m\pm1)$ matrix satisfying the following
conditions:
\eqn\ospelement{\eqalign{
M_{i,i+2k}&=(-)^{k+1}M_{p+1-i-2k,p+1-i}\qquad\qquad\cr
M_{i,i+2k+1}&=(-)^{k+1}M_{p-i-2k,p+1-i}\qquad\qquad
           ~~~\cr
(\ p=4m\pm1 & \quad ; \quad  i=1,\dots , p  \quad ; \quad
k=0, \pm 1,\dots \ ) \, .}}
The principal embedding of $osp(1|2)_{{\rm pal}} \subset osp(2m\pm1|2m)$
is again given by
$$
J_-=\sum _i f_i \quad , \quad J_+=\sum_{ij}a^{ij}e_i \quad , \quad
H=\{J_+,J_-\} \, .
$$
The bases \chevalleyp~and \chevalleym~
have been chosen such as to lead
to the same explicit expressions for $J_{\pm}$ and $H$ as in
the $sl(n+1|n)$ case, see
eq.\sppal. The highest weight generators are also the same,
except that we have to exclude
those $M_k$ which do not belong to
$osp(2m\pm1|2m)$. This leaves us with $M_{2,3\,({\rm mod}\,4)}$.
The $\th$-component of an
$osp(2m\pm1|2m)$-valued connection in highest weight gauge
is thus given by
\eqn\ospcon{-{\cal A}_{N=1}=-J_-+\sum_{k=2,3\,{\rm mod}\,4}
            V_{k+1}M_k \, .}
The number of fields occuring in ${\cal A}_{N=1}$
equals the rank of the corresponding algebra,
namely $2m$ for $osp(2m+1|2m)$ and $2m-1$ for $osp(2m-1|2m)$.
Note that in contrast to the $sl(n+1|n)$ case there
is no superfield $V_2$ of weight one. We are thus dealing
with higher spin extensions of the $N=1$ superconformal
algebra. The $(2n+1)$-th order operator associated to the
corresponding
first order system is symmetric in the sense of ref.\cuba~
(thus allowing for the factorization considered in eq.(37)
below).

\goodbreak
\noindent {\bf 6. An example: $osp(3|2)$}

\noindent{\bf General set-up}:
The simplest $N=1$ SWA occurs for the algebra $osp(1|2)$;
in this case the zero-curvature condition \zerocurv~ yields
the $N=1$ superconformal Ward identity which
is easily integrated to give the $N=1$ superconformal
algebra \dg. Instead of this very simple example, we will rather
consider the algebra $osp(3|2)\subset sl(3|2)$ to illustrate
the procedure outlined above. For $sl(3|2)$, the operator \hwg~
takes the form
\eqn\deltaex{
\nabla=\pmatrix{D&2V_2&2V_3&V_4&V_5\cr
                -1&D&V_2&V_3&V_4\cr
                  &-1&D&V_2&2V_3\cr
                     &&-1&D&2V_2\cr
                     &&&-1&D\cr}\, ,}
where $V_k$ is even (odd) for $k$ even (odd) \gt.
For the matrix which generates the transformation
between $\nabla$ and $\nabla^\prime$ one then finds
\eqn\xex{
X=\pmatrix{1&0&4V_2&DV_2-V_3&2\p V_2+2V_2^2+2DV_3 +V_4 \cr
            &1&0&3V_2&2DV_2+2V_3\cr
            & &1&0&2V_2\cr
            & & &1&0\cr
            & & & &1\cr}}
and for the relations between the $a_k$ and the
$V_k$ one gets the expressions
\eqn\relav{
\eqalign{a_2&=6V_2\cr
         a_3&=3V_3+3DV_2\cr
         a_4&=2V_4+DV_3+3\p V_2+8V_2^2\cr
         a_5&=V_5+DV_4+2\p V_3+2 D^3 V_2
                 +8V_2DV_2+8V_2 V_3 \, .}}

So far we have still looked at the graded algebra
$sl(3|2)$ which would lead
to the $N=2$ $W_3$-algebra of ref.\hn.
To go over to the $osp(3|2)$ case,
we now set $V_2=V_5\equiv 0$; then we have
\eqn\set{
\nabla=D-J_-+M_2 V_3+M_3 V_4 \, .}
For $\bar\A$ we make the
most general Ansatz compatible with the $osp(3|2)$ structure:
\eqn\abar{\bar\A=\pmatrix{\alpha_1&a_1&\beta_1&b&0\cr
                          c_1&\alpha_2&a_2&\beta_2&b\cr
                          \gamma_1&c_2&0&-a_2&\beta_1\cr
                          d&\gamma_2&c_2&-\alpha_2&-a_1\cr
                          0&-d&\gamma_1&c_1&-\alpha_1\cr}\, . }
Here, small Latin (Greek) letters denote Grassmann even (odd)
superfields (cf. eq.\decomp).
In eq.\abar~ we have suppressed the indices of all matrix
elements. These can be determined by
going back to eq.\f~ and noting that
$f \equiv f^z \in {\cal F}_{-2}$ in the present case. This fixes the
indices of all elements of $\vec F ^{\prime}$. By virtue of the relation
$\vec F = X \vec F^{\prime}$ the components of $\vec F$ then have the same
index structure as those of $\vec F ^{\prime}$.
In the system of equations
$0 = \bar{\nabla} \vec F = (\dab - \bar{\A} ) \vec F$, we encounter
the derivatives $\dab - \alpha_1$ and
$\dab  - \alpha_2$ so that
$\alpha_1 = (\alpha_1)_{\tb}$ and
$\alpha_2 = (\alpha_2)_{\tb}$. This determines the index structure of
all other elements of $\bar\A$:
$$
b_{\tb \th z} \quad , \quad (\beta_i)_{\tb z} \quad , \quad
(a_i)_{\tb \th} \quad , \quad
(\alpha_i)_{\tb} \quad , \quad
(c_i)_{\tb}^{\ \th} \quad , \quad
(\gamma_i)_{\tb}^{\ z} \quad , \quad
d_{\tb}^{\ \th z} \quad  \quad  (i = 1,2 ) \, .
$$

The transformation laws of these variables under a superconformal
change of coordinates,
$(z, \zb , \th , \tb ) \to
(\tilde{z}, \tilde{\zb} , \tilde{\th} , \tilde{\tb} )$, are
derived along the same lines from the one of
$f \in {\cal F}_{-2}$, i.e. from
$$
f \to e^{-2w} f \qquad {\rm where} \quad e^{-w} \equiv D\tilde{\th} \,.
$$ From
this transformation law we first deduce the one of $\vec F ^{\prime}$
which fixes the one of $\vec F$ by virtue of the relation
$\vec F = X \vec F^{\prime}$
and of the transformation laws of the $V_k$. One finds that
$\vec F \to S \vec F$ with
$$
S =
\pmatrix{e^{2w} & 2e^{2w} Dw & -2e^{2w}\p w & -2e^{2w} (Dw)\p w
                                           &  2e^{2w}(\p w)^2 \cr
         0 & e^{w} & -e^{w} Dw & -e^{w}\p w & 2e^{w} (Dw)\p w \cr
         0 & 0  & 1  & Dw & -2\p w \cr
         0 & 0 & 0 & e^{-w} & -2e^{-w} Dw \cr
         0 & 0  & 0 &  0  & e^{-2w} \cr } \, .
$$
Since $\dab w =0$, we have
$(\dab \vec F) \to e^{\bar{w}} \widehat S (\dab \vec F)$
and therefore the condition
$\bar{\nabla} \vec F = 0$ is superconformally covariant if and only if
the matrix $\bar{\A}$ transforms according to
\eqn\transf{
\bar\A \to e^{\bar{w}} \widehat{S} \bar\A S^{-1} \, .}
The elements of $\bar\A$ which have the simplest transformation laws are
easily seen to be $d$ and $\gamma_1$:
\eqn\simple{
\eqalign{d& \to e^{\bar{w}} e^{-3w} d \cr
\gamma_1& \to e^{\bar{w}} e^{-2w} \left[ \gamma_1 - (Dw) d \right] \,.}}
Henceforth,
\eqn\combin{
\left[ \gamma_1 - {1\over 3} Dd \right] \to e^{\bar{w}} e^{-2w} \left[ \gamma_1
-{1 \over 3} Dd \right] }
and we conclude that both $d$ and
$\gamma_1 - {1\over 3}Dd$ transform like superconformal fields.

\noindent{\bf Ward identities}:
Substituting ${\cal A}$ and $\bar{\cal A}$ (as given by eqs.\set~ and
\abar, respectively) into the
zero-curvature condition \zerocurv, we obtain a set of
coupled equations for $V_3, V_4$ and the coefficients of $\bar{\cal A}$.
Allmost all of these can be solved algebraically for
the coefficients of $\bar\A$ in terms of $V_3, \, V_4$
and $\gamma_1, \, d$; one is only left with two differential
equations which are in fact the
Ward identities for the
underlying SWA.
In terms of the superconformal fields
$H_3 \equiv \gamma_1-{1\over 3}Dd$ and
$H_4 \equiv {4\over 3} d$, these identities take
the form
\eqn\ward{\eqalign{
\Bigl\{\bar D&-H_3\p
+\half(DH_3)D
-{3\over2}(\p H_3)\Bigr\}V_3\cr
&=-\half D^5 H_3+
\Bigl\{{3\over2}H_4\p
+\half(DH_4)D
+2(\p H_4)\Bigr\}V_4\cr
\noalign{\vskip.5cm}
\Bigl\{\bar D&-H_3\p
+\half(DH_3)D
-2(\p H_3)
-{3\over4}H_4 D^3
-{5\over4}(DH_4)\p
-{5\over4}(\p H_4)D
-{5\over2}(D^3 H_4)
\Bigr\}V_4\cr
&={1\over 8}D^7 H_4+\Bigl\{{3\over8}H_4\p^2+{1\over 4}(DH_4)D^3
+(\p H_4)\p+\half(D^3 H_4)D+{3\over 4}(\p^2 H_4)\Bigr\}V_3\cr
&\quad+{9\over 8}V_3(DV_3)H_4+{9\over 2}V_3 V_4 H_4 \, . }}
In the remainder of this section, we will explore these equations
and the information they contain. First, we note that the
superfields involved in eq.\ward~ have
the index structure
$(H_3)_{\bar{\theta}}^{\ z}, (H_4)_{\bar{\theta}}^{\ \theta z}$
and $(V_3)_{\theta z}, (V_4)_{zz}$, respectively. More specifically,
$(H_3)_{\tb}^{\ z}$ can be identified with the super Beltrami coefficient
$H_{\bar{\theta}}^{\ z}$ parametrizing superconformal classes
of metrics \dg~ and
$(V_3)_{\theta z}$ is a multiple of the superstress tensor;
this correspondence is confirmed by the fact that the previous set of
identities reduces to the $N=1$
superconformal Ward identity for
$H_4 = 0 = V_4$.
(The latter is the anomalous Ward identity expressing the
superdiffeomorphism invariance
of the generating functional $Z_c[H_3]$ in superconformal
field theory \dg.)

\noindent{\bf Operator product expansions}:
To derive the singular parts of the operator products from the
Ward identities \ward, we follow reference \dg~
and we consider them
at ${\cal Z}_1=(z_1,\bar z_1,\theta_1,\bar\theta_1)$, then multiply by
$\theta_{12}/ {\cal Z}_{12}$
(with $\theta_{12}= \theta_1 -\theta_2$ and ${\cal Z}_{12} = z_1 - z_2 -
\theta_1 \theta_2$) and subsequently integrate with respect to
${\cal Z}_1$. Thereafter, we
replace $V_3$ by $\delta Z_c / \delta H_3$ and
$V_4$ by $\delta Z_c / \delta H_4$
where $Z_c[H_3 , H_4]$ denotes
a generating functional.
Application of the functional derivatives
$\delta / \delta H_3,
\delta / \delta H_4$ to the so-obtained equations at the point $H_3 =0=H_4$
then leads to the
operator products. If we define the correspondence
$\delta / \delta H_3 \sim T$, where $T$ is the
superstress tensor and
$\delta / \delta H_4 \sim W_4$, where
$W_4$ is a superconformal field of weight ${2}$, we find the following
expressions for the singular parts of the operator products:
\def\zz{{\cal Z}}
\eqn\ope{\eqalign{
T({\cal Z}_1)T({\cal Z}_2)&={1\over\zz_{12}^3}+ \Bigl[ {3\over 2}
{\theta_{12}\over\zz_{12}^2}+\half{1\over\zz_{12}}D_2
+{\theta_{12}\over\zz_{12}}\p_2 \Bigr]  T(\zz_2)\cr
T(\zz_1)W_4(\zz_2)&= \Bigl[ 2{\theta_{12}\over\zz_{12}^2}
+\half{1\over\zz_{12}}D_2 +{\theta_{12}\over\zz_{12}}
\p_2 \Bigr]  W_4(\zz_2)\cr
-W_4(\zz_1)W_4(\zz_2)=&{3\over 4}{1\over\zz_{12}^4}+
\Bigl[ {5\over 2}{1\over\zz_{12}^2}
+{5\over 4}{\theta_{12}\over\zz_{12}^2}D_2
+{5\over 4}{1\over\zz_{12}}\p_2
+{3\over4}
{\theta_{12}\over\zz_{12}}D_2 ^3\Bigr] W_4(\zz_2)\cr
&+\Bigl[ {3\over 2}{\theta_{12}\over\zz_{12}^3}+\half{1\over\zz_{12}^2}
D_2+{\theta_{12}\over\zz_{12}^2}\p_2 +{1\over4}{1\over\zz_{12}}D_2 ^3
+{3\over8}{\theta_{12}\over\zz_{12}}\p_2 ^2\Bigr] T(\zz_2)\cr
&+{9\over8}{\theta_{12}\over\zz_{12}}T(\zz_2) \Bigl[ D_2T(\zz_2)
+4 W_4(\zz_2)\Bigr]  \, . }}
The product of $T$ with $W_4$ expresses the fact that $W_4$
is a primary field of weight $2$.
The terms in the operator products which involve products of fields
have to be regularized in the quantum version of the algebra
\footnote{$^*$}{The quantum algebra has been computed
for this case in ref.\blum~
by solving the Jacobi identities.}.

\noindent{\bf Relation with Poisson Algebra}:
Let us now go back to the identities \ward~ and rewrite them in the
form
\eqn\warda{\bar D V_i= \sum_{j=3,4}(-)^i L^{ij}H_j \qquad \quad
(\, i = 3,4 \, )}
with
\eqn\final{
\eqalign{
L^{33}=&\; \half\Bigl[ D^5+3V_3\partial
+(DV_3)D+2(\partial V_3)\Bigr]\cr
L^{34}=&\; -\Bigl[ 2V_4\partial-{1\over2}(DV_4)D
+{3\over2}(\partial V_4) \Bigr] \qquad , \qquad
L^{43}=2V_4\partial-{1\over2}(DV_4)D
+(\partial V_4)\cr
L^{44}=&\ {1\over 8}\Bigl[ D^7+6V_3\p^2+4(DV_3)D^3+8(\p V_3)\p
+2(D^3 V_3)D+3(\p^2 V_3)+{9} V_3(DV_3)\Bigr]\cr
& +{5\over 2}\Bigl[V_4 D^3+\half(D V_4)\p
+\half (\p V_4)D+{3\over10}(D^3 V_4)+{9\over 5} V_3 V_4
\Bigr] \, .} }
Then, the different
operators $L^{ij}$ are just the operators defining the Poisson brackets
between the fields $V_i$,
\eqn\poisson{\{V_i({\cal Z}_1),V_j({\cal Z}_2)\}=L^{ij}_{{\cal Z}_1}
\delta({\cal Z}_1-{\cal Z}_2)\,,}
where $\delta(\Z_1-\Z_2)=(\theta_1-\theta_2)\delta(z_1-z_2)$.
The equivalence between the Poisson algebra
\poisson~and the operator product algebra \ope~ follows easily from the
correspondence
\eqn\corres{
{\theta_{12}\over\Z_{12}^{r+1}}\,\leftrightarrow\,(-)^r
{1\over r!}\partial_1^{r}\delta(\Z_1-\Z_2)\,\,,\qquad
{1\over\Z_{12}^{r+1}}\,\leftrightarrow\,(-)^r{1\over r!}
D_1^{2r+1}\delta(\Z_1-\Z_2)\,.}

As an
independent check of the algebra \poisson~
we have explicitly derived these brackets by using
the second Gelfand-Dickey
Hamiltonian structure.
For that purpose it is convenient
to take the basic differential operator in the factorized form \cuba,
\eqn\fac{
{\cal L}^{(n)} =(D+\phi_1)(D+\phi_2)D(D-\phi_2)(D-\phi_1) \, ,}
where the fields $\phi_i$ are related to $V_3, V_4$
by the generalized Miura transformation
\eqn\miura{\eqalign{
V_3&=-{2\over3}\partial\phi_1-{1\over3}\phi_1 D\phi_1
+{1\over3}\partial\phi_2+{1\over3}\phi_2 D\phi_2\cr
V_4&={1\over3}D^3\phi_1+{1\over3}\phi_1\partial\phi_1
     +{1\over6}(D\phi_1)^2-{2\over3}D^3\phi_2\cr
   &\qquad+{2\over3}\phi_2\partial\phi_2-{2\over3}(D\phi_2)^2
          -\phi_1\partial\phi_2-\phi_1\phi_2 D\phi_2 \, . }}
By virtue of the
supersymmetric version of the Kupershmidt-Wilson
theorem,
the second Gelfand-Dickey bracket for the fields $V_i$ is then given
by the first bracket for the fields $\phi_i$, which reads \cuba
\eqn\gelfand{
\lbrace\phi_i({\cal Z}_1),\phi_j({\cal Z}_2)\rbrace
=(-)^i\delta_{ij}D_{{\cal Z}_1}
\delta({\cal Z}_1 -{\cal Z}_2) \, .}
After a considerable amount of work and rescaling the
fields
we have explicitly derived eq.\poisson.
Concerning the derivation we remark that
the following general
result is helpful: if the fundamental Poisson bracket between
the fields $u_i$ with Grassmann parity $|u_i|=(-)^i$
is given by
$$
\{u_i(\Z_1),u_j(\Z_2)\}=L^{ij}_{{\cal Z}_1}\delta(\Z_1-\Z_2) \, ,
$$
then the
bracket of two differential polynomials of $u_i$ is given by
$$
\eqalign{\{{\cal F}(\Z_1),{\cal H}(\Z_2)\}=&
\sum_{i,j}\sum_{p,q=0}^\infty
(-)^{|{\cal H}|+(|{\cal F}|+i+q)(i+q)+\half p(p+1)
+j(p+1)}{\partial{\cal F}\over\partial u_i^{(q)}}
(\Z_1)D_{{\cal Z}_1}^q\circ\cr
&\qquad\qquad L_{{\cal Z}_1}^{ij}\circ D_{{\cal Z}_1}^p\circ
{\partial{\cal H}\over\partial u_j^{(p)}}(\Z_1)\delta(\Z_1-\Z_2) \, ,}
$$
where $u_i^{(p)}=D^p u_i$.

\noindent{\bf Covariance}: If written in the form \warda~, the Ward identities
are manifestly covariant
since the
operators $L^{ij}$ represent
linear combinations of superconformally covariant operators of the types
constructed in refs.\cco~ and \gt.
In fact, if
we identify the superstress tensor $V_3$ with a superprojective
connection ${\cal R}$, eqs.\warda~ take the compact form
\eqn\covar{\eqalign{
\dab V_3 & = \ - \half {\cal L}_2 H_3 +6J_{4,-3}^2(V_4,\cdot) H_4  \cr
\dab V_4 & = \  \Bigl[ {1\over 8} {\cal L}_3
+ {5\over 2} M^{(3)}_{V_4} \Bigr] H_4
+4 J^2 _{4,-2} (V_4 , \cdot ) H_3 \, .}}
Here, ${\cal L}_2$ and ${\cal L}_3$ are super Bol operators (depending
only on the projective connection $V_3 = {\cal R}$) while
$M^{(3)}_{V_4}$ and $J^2 _{4,k} (V_4 , \cdot )$ are linear operators
depending on both $V_3$ and the conformal field $V_4$.
These operators are superconformally covariant when acting on
$H_3 \in \F_{-2} \otimes \bar{\F}_1$ and
$H_4 \in \F_{-3} \otimes \bar{\F}_1$, respectively.

\noindent{\bf Generalization to $osp(2m\pm 1|2m)$}:
Quite generally, for
$W_n$ to be a primary field of weight ${n\over2}$, the
Ward identities for the fields $V_3$ and $V_n$ need to have the
covariant form
\eqn\wardn{\eqalign{
&\Bigl\{\bar D-H_3\p
+\half(DH_3)D
-{3\over2}(\p H_3)\Bigr\}V_3\cr
& \qquad =-\half D^5 H_3+
\Bigl\{{n-1 \over2}H_n\p
+\half (-)^n (DH_n)D +{n \over 2}
(\p H_n)\Bigr\}V_n\cr
\noalign{\vskip.5cm}
&\Bigl\{\bar D-H_3\p
+\half(DH_3)D
-{n \over 2} (\p H_3)
\Bigr\}V_n
= {\rm terms\,\,involving\,\,}H_{n\neq3} \,.}}
This corresponds to the operator products
\eqn\ope{\eqalign{
T({\cal Z}_1)T({\cal Z}_2)&={1\over\zz_{12}^3}+ \Bigl[ {3\over 2}
{\theta_{12}\over\zz_{12}^2}+\half{1\over\zz_{12}}D_2
+{\theta_{12}\over\zz_{12}}\p_2 \Bigr]  T(\zz_2)\cr
T(\zz_1)W_n(\zz_2)&= \Bigl[ {n \over 2} {\theta_{12}\over\zz_{12}^2}
+\half{1\over\zz_{12}}D_2 +{\theta_{12}\over\zz_{12}}
\p_2 \Bigr]  W_n(\zz_2)\,,}}
or, equivalently, to Poisson brackets involving
\eqn\brack{
L^{3n}=(-)^{n+1}\left[ {n\over 2}V_n\partial-\half(-)^n(DV_n)D
+{n-1\over2}(\partial V_n)\right]
\qquad {\rm for} \ n\neq3}
and $L^{33}$ as in eq.\final.

\noindent {\bf 7. Concluding remarks}

To summarize, we have provided a general algorithm for deriving the
classical SWA's associated to the super Lie algebras
$sl(n+1|n)$ and $osp(2m\pm1|2m)$.
This was explicitly
demonstrated with the $osp(3|2)$ example. As mentionned above,
a similar calculation for the simple case
of $osp(1|2)$ leads to the super Virasoro
algebra or, equivalently, to the second Hamiltonian
structure for the super KdV equation. The Poisson structure derived here
for the $osp(3|2)$ case should correspond to another
supersymmetric integrable hierachy,
associated with the Lax operator $D{\cal L}^{(2)}$.

\bigskip
\bigskip
\noindent {\bf Acknowledgments}

F.G. wishes to thank F.Delduc for a helpful remark on super algebras.
\bigskip
\bigskip

\noindent {\bf Appendix A: Gauge theories with supergroups}

In this appendix we work out the superspace
formulation of gauge theories based on
supergroups \drs~ and we discuss the integrability conditions
for gauge covariant differential equations.
Upon restriction to the even part
of the supergroup, all formulae reduce to the standard ones
for ordinary groups \gates~.

For a super Riemann surface,
the canonical basis of the tangent space
is given by the vector fields
$(D_A) = (\p , \pab , D, \dab)$. The associated Lie brackets read
$$
[ D_A , D_B \} \equiv D_A D_B - (-)^{ab} D_B D_A
= -T_{AB}^{\ \ \ C} D_C \, .
$$
Here, the anholonomy coefficients $T_{AB}^{\ \ \ C}$
correspond to the rigid superspace torsion and the only
non-vanishing coefficients are given by
$T_{\th \th} ^{\ \ z} = -2 = T_{\tb \tb}^{\ \ \zb}$.
The graded commutator of the gauge covariant derivatives
$\nabla_A = D_A - {\A}_A$ defines the components $\F_{AB}$
of the curvature associated to the connection $\A$:
\eqn\graded{\eqalign{
[ \nabla_A , \nabla_B \} & = \ - \F_{AB} - T_{AB}^{\ \ \ C} \nabla_C  \cr
\F_{AB} & = \ D_A \A_B - (-)^{ab} D_B \A_A -
[ \A_A , \A_B \} + T_{AB}^{\ \ \ C} \A_C \, .}}
If the connection takes its values in a super Lie algebra,
the graded commutators of the connection components are defined by the
relations
\eqn\commut{\eqalign{
{\rm even}/{\rm even} :& \qquad
[ \A_z , \A_{\zb} \} = \A_z \A_{\zb} - \A_{\zb} \A_z    \cr
{\rm odd}/{\rm odd } :& \qquad
[ \A_{\th} , \A_{\tb} \} =
\widehat{\A} _{\th} \A_{\tb} + \widehat{\A}_{\tb}
\A_{\th} \cr
{\rm even}/{\rm odd } :& \qquad
[ \A_z , \A_{\tb} \} =  \widehat{\A} _z \A_{\tb} - \A_{\tb} A_z
= -
[ \A_{\tb} , \A_z \}}}
and analogously for $\A_z \leftrightarrow \A_{\zb}, \,
A_{\th} \leftrightarrow \A_{\tb}$ (as well as for the commutators of
covariant derivatives).
In order to obtain these
graded commutators in a systematic way, it is convenient to  multiply the
odd superfields $\A_{\th}, \, \A_{\tb}$ by Grassmann numbers
$\varepsilon, \, \varepsilon^{\prime}$ so as to obtain
even superfields:
the relations \commut~ then follow from the ordinary commutators
\eqn\ordinary{\eqalign{
[ \varepsilon \A_{\th} , \varepsilon^{\prime} \A_{\tb} ] & = \
(\varepsilon \A _{\th})( \varepsilon^{\prime} \A_{\tb}) -
(\varepsilon^{\prime} \A _{\tb})( \varepsilon \A_{\th})
\equiv - \varepsilon \varepsilon^{\prime}
[ \A_{\th} , \A_{\tb} \}  \cr
[ \A_z , \varepsilon \A_{\tb} ] & = \
\A _z ( \varepsilon \A_{\tb} ) -
( \varepsilon \A _{\tb} ) \A_z
\equiv \varepsilon
[ \A_z , \A_{\tb} \} }}
and the relations
\eqn\signs{
\A_{\th} \varepsilon = - \varepsilon \widehat{\A} _{\th}
\qquad , \qquad
\A_z \varepsilon =  \varepsilon \widehat{\A} _z \, . }

If the vector $\vec F$ satisfies the covariant differential equations
\eqn\grad{
\nabla_{\th} \vec F =0 \qquad {\rm and} \qquad
\nabla_{\tb} \vec F =0 \, ,}
then the relations \graded~ imply
the integrability condition
$ 0=[\nabla_{\th} , \nabla_{\tb} \} \vec F = - \F_{\th \tb} \vec F$,
i.e.
\eqn\ff{
{\cal F}_{\th \tb} =0 \, . }
This is just our zero-curvature equation \zerocurv.
Furthermore, eqs.\graded~ imply the integrability condition
$$
0 = [ \nabla_{\th} , \nabla_{\th} \} \vec F =
- {\cal F}_{\th \th} \vec F + 2 \nabla_z \vec F
$$
and analogously for
$[ \nabla_{\tb} , \nabla_{\tb} \} \vec F$, i.e.
one has to require that
\eqn\integ{
\nabla_z \vec F = {1\over 2} {\cal F}_{\th \th} \vec F
\qquad , \qquad
\nabla_{\zb} \vec F = {1\over 2} {\cal F}_{\tb \tb} \vec F\, .}

If we now impose the redefinition constraints
$\F_{\th \th} =0$ and
$\F_{\tb \tb} =0$, we can use the definition \graded~ to
express the spatial components
$\A_z$ and $\A_{\zb}$ in terms
of $\A_{\th}$ and $\A_{\tb}$, respectively:
\eqn\redefine{
\A_z = D\A_{\th} - \widehat{\A} _{\th} \A_{\th}  \qquad , \qquad
\A_{\zb} = \dab \A_{\tb} - \widehat{\A} _{\tb} \A_{\tb} \, .}
In this case, eqs.\integ~ reduce to
$\nabla_z \vec F =0$ and
$\nabla_{\zb} \vec F =0$ and
compatibility of these equations
with each other and with eqs.\grad~
requires that all remaining components of
the curvature vanish:
\eqn\zero{
0=\F_{z \zb }= \F_{z \th}
= \F_{z \tb}
=\F_{\zb \th } = \F_{\zb \tb} \, .}

\vskip.5cm
\noindent {\bf Appendix B: $osp(2m\pm1|2m)$ in diagonal grading}

One usually defines \corn~
the superalgebra $osp(2m\pm1|2m)$ as the set of
supermatrices $M=\smat{A}{B}{C}{D}$ which
satisfy $M^{sT}G+GM=0$ where the supertranspose of $M$ is defined by
$M^{sT}=\smat{A^T}{C^T}{-B^T}{D^T}$ and an invariant metric by
$G=G_{so(2m\pm1)}\oplus G_{sp(2m)}$. For the
invariant $so(2m\pm1)$
and $sp(2m)$ metrics we choose, respectively,
$$
G_{so(2m\pm1)}=
\pmatrix{&&&1\cr
         &&\cdot & \cr
         &\cdot &&  \cr
         1&&&}
\qquad , \quad
G_{sp(2m)}=\pmatrix{0&-{\bf 1}_m\cr{\bf 1}_m&\phantom{-}0}\, .
$$

Let us now discuss the two cases $osp(2m-1|2m)$ and
$osp(2m+1|2m)$ in turn.
\vskip.3cm

\noindent $\underline{osp(2m-1|2m)}$:
\noindent There are $4m^2-2m+1$ even and $2m(2m-1)$ odd generators.
The
Chevalley basis in block grading is represented by the elements
\eqn\bc{
\eqalign{e_{2i-1}&=E_{m+1-i,4m-i}+E_{3m-i,m+1-i}
\quad\qquad (\, i=1,\dots,m \, )\cr
e_{2i}&=E_{m+i,3m-i}-E_{4m-i,m-i}
\quad\qquad\qquad (\, i=1,\dots,m-1 \, ) \cr
\noalign{\vskip.3cm}
f_{2i-1}&=E_{m-1+i,3m-i}-E_{4m-i,m+1-i}\cr
f_{2i}&=-E_{m-i,4m-i}-E_{3m-i,m+i}\cr
\noalign{\vskip.3cm}
h_{2i-1}&=E_{3m-i,3m-i}-E_{4m-i,4m-i}-E_{m+1-i,m+1-i}+E_{m-1+i,m-1+i}\cr
h_{2i}&=E_{m-i,m-i}-E_{m+i,m+i}-E_{3m-i,3m-i}+E_{4m-i,4m-i}\, .}}
\vskip.3cm
The transition between matrices in block and diagonal grading is mediated
by a similarity transformation,
$M_{\rm diag.}=L^{-1}M_{\rm block}L$.
The choice of $L$ that leads to the expression for the
generators of $osp(1|2)_{\rm pal}$ given by
eq.\sppal~ is
\eqn\sim{
\eqalign{
L&=\sum_{i=0}^{m-1}(-)^{i+1}E_{2m+i,2i+1}+\sum_{i=1}^m(-)^i
E_{2m-i,2i}\cr
&\qquad\qquad+(-1)^{m+1}\sum_{i=1}^n E_{4m-i,2m+2i-1}
+(-)^m\sum_{i=1}^{m-1}E_{m-i,2m+2i}\, .}}
The invariant metric then becomes
\eqn\inv{
G=\sum_{i=1}^{4m-1}(-)^{m+\lbrack{i+1\over2}\rbrack}
E_{i,4m-i} \, ,}
where $\lbrack i\rbrack$ denotes the integer part. For the elements
of $M^{sT}$ we find
\eqn\supertrans{\eqalign{
\left(M^{sT}\right)_{i,i+2k}&=M_{i+2k,i} \qquad  \quad
(\  i=1,\dots , 4m-1  \ \ ; \ \
k=0, \pm 1,\dots \ ) \cr
\left(M^{sT}\right)_{i,i+2k+1}&=(-)^i M_{i+2k+1,i} \, .}}
The Chevalley generators and the general matrix element
in diagonal grading
are also obtained easily and are as
given in eqs.\chevalleym~and \ospelement.
\vskip.4cm

\noindent$\underline{osp(2m+1|2m)}$:
\noindent There are $2m(2m+1)$ even and odd generators.
The Chevalley basis in block grading is
($i=1,\dots,m$)
\eqn\horse{
\eqalign{e_{2i-1}&=E_{m+2-i,4m+2-i}+E_{3m+2-i,m+i}\cr
e_{2i}&=E_{m+1+i,3m+2-i}-E_{4m+2-i,m+1-i}\cr
\noalign{\vskip.3cm}
f_{2i-1}&=E_{m+i,3m+2-i}-E_{4m+2-i,m+2-i}\cr
f_{2i}&=-E_{m+1-i,4m+2-i}-E_{3m+2-i,m+1+i}\cr
\noalign{\vskip.3cm}
h_{2i-1}&=E_{3m+2-i,3m+2-i}-E_{4m+2-i,4m+2-i}
-E_{m+2-i,m+2-i}+E_{m+i,m+i}\cr
h_{2i}&=E_{m+1-i,m+1-i}-E_{m+1+i,m+1+i}
-E_{3m+2-i,3m+2-i}+E_{4m+2-i,4m+2-i }\, .}}
\vskip.3cm
\noindent For the matrix $L$ we find
\eqn\simt{
\eqalign{
L&=\sum_{i=0}^{m}(-)^{i}E_{2m+1-i,2i+1}+\sum_{i=1}^m(-)^i
E_{2m+1+i,2i}\cr
&\qquad\qquad+(-1)^{n+1}\sum_{i=1}^m E_{4m+2-i,2m+2i}
+(-)^m\sum_{i=1}^{m}E_{m+1-i,2m+1+2i}\, .}}
The invariant metric then becomes
\eqn\metr{
G=\sum_{i=1}^{4m+1}(-)^{m+\lbrack{i\over2}\rbrack}
E_{i,4m+2-i} }
and the elements of $M^{sT}$ take the form
\eqn\supertrans{\eqalign{
\left(M^{sT}\right)_{i,i+2k}&=M_{i+2k,i} \qquad  \quad
(\  i=1,\dots , 4m+1  \ \ ; \ \
k=0, \pm 1,\dots \ ) \cr
\left(M^{sT}\right)_{i,i+2k+1}&=(-)^{i+1} M_{i+2k+1,i} \, . }}
For the Chevalley generators and the general matrix element
in diagonal grading
we refer again to the main body of the text
(eqs.\chevalleyp~and \ospelement).

\listrefs
\bye